\documentclass[12pt]{article}
\usepackage{epsf}

\def\Tr                 {\mathop{\rm Tr}}
\def\none               {\multicolumn{2}{c|}{---}}
\newcommand{\be}{\begin{equation}}
\newcommand{\ee}{\end{equation}}
\newcommand{\beq}{\begin{equation}}
\newcommand{\eeq}{\end{equation}}
\newcommand{\bea}{\begin{eqnarray}}
\newcommand{\eea}{\end{eqnarray}}
\def\lsi{\raise0.3ex\hbox{$<$\kern-0.75em\raise-1.1ex\hbox{$\sim$}}}
\def\gsi{\raise0.3ex\hbox{$>$\kern-0.75em\raise-1.1ex\hbox{$\sim$}}}

\newcommand{\gsim}{\mathbin{\gsi}}
\newcommand{\MeV}{\mathop{\rm MeV}}
\newcommand{\GeV}{\mathop{\rm GeV}}
\newcommand{\fm}{\mathop{\rm fm}}

\begin{document}

\begin{titlepage}

\noindent

\begin{flushright}
DAMTP-2001-69 \\
OUTP-01-44P\\
hep-lat/0108022\\
\hspace{1em}\\
August 2001
\end{flushright} \begin{centering} \vfill

{\bf ON THE GLUEBALL SPECTRUM IN \\
\boldmath{${\cal O}(a)$}--IMPROVED LATTICE QCD}
\vspace{0.8cm}

{\sl UKQCD Collaboration}

\vspace{0.8cm}

A. Hart$^{\rm a}$ and 
M. Teper$^{\rm b}$

\vspace{0.3cm}
{\em $^{\rm a}$%
DAMTP, CMS, University of Cambridge, Wilberforce Road, \\
Cambridge CB3 0WA, England\\}
\vspace{0.3cm}
{\em $^{\rm b}$%
Theoretical Physics, University of Oxford, 1 Keble Road, \\
Oxford OX1 3NP, England\\}

\vspace{0.7cm}
{\bf Abstract.}
\end{centering}

\vspace{1ex}
\noindent
We calculate the light `glueball' mass spectrum in $N_f=2$ lattice QCD
using a fermion action that is non--perturbatively ${\cal O}(a)$
improved. We work at lattice spacings $a \sim 0.1 \fm$ and with quark
masses that range down to about half the strange quark mass. We find
the statistical errors to be moderate and under control on relatively
small ensembles. We compare our mass spectrum to that of quenched QCD
at the same value of $a$.  Whilst the tensor mass is the same (within
errors), the scalar mass is significantly smaller in the dynamical
lattice theory, by a factor of $\sim 0.84\pm 0.03$.  We discuss what
the observed $m_q$ dependence of this suppression tells us about the
dynamics of glueballs in QCD. We also calculate the masses of flux
tubes that wind around the spatial torus, and extract the string
tension from these. As we decrease the quark mass we see a small but
growing vacuum expectation value for the corresponding flux tube
operators. This provides clear evidence for `string breaking' and for
the (expected) breaking of the associated gauge centre symmetry by sea
quarks.

\vfill


\end{titlepage}

\section{Introduction}

The spectrum of the lightest states in the SU(3)
gauge theory is by now well established (for a review see
\cite{teper98} 
and for more recent work see
\cite{morningstar99,lucini01}). 
The lightest state is a scalar ($J^{PC}=0^{++}$) with mass around
$3.6\sqrt K$, followed by a tensor ($J^{PC}=2^{++}$) at around
$5.1\sqrt K$, with $K$ being the string tension. We may also regard
this as being the glueball spectrum of the quenched approximation to
QCD. In QCD the string tension is believed to have a value $\surd K
\simeq 450 \MeV$, so this means that the lightest scalar and tensor in
(quenched) QCD are expected to have masses of approximately $1600
\MeV$ and $2300 \MeV$ respectively.

These masses have been arrived at by performing simulations 
for a range of lattice spacings, and then removing the 
discretisation effects by making an extrapolation to the 
continuum limit. To obtain the masses in physical $\MeV$ units
involves the use of a physical scale parameter, most commonly 
either the string tension (e.g. in
\cite{lucini01}) 
or the intermediate Sommer scale, $r_0$
\cite{sommer94} 
(e.g. in
\cite{morningstar99}). 
The variations in the continuum mass estimates that arise from 
the use of different scale parameters are small but
significant, and this highlights 
the inevitable uncertainties that exist in using the computationally 
attractive quenched approximation to describe the real world.

In the case of glueballs, the real world appears to be considerably 
more complicated than quenched QCD. In the mass region around
$1500 \MeV$ there are at least two and probably three
flavour singlet scalar resonances.  While the precise nature of 
these states is unclear, an attractive interpretation is that they 
are the result of mixing between the two flavour singlet pure 
$q\bar{q}$ states (`quarkonia') of the scalar nonet, and
the lightest scalar glueball whose mass, as we have remarked above,
should be in this mass range. The mixing models
(see e.g.
\cite{lee99,close01}) 
are not sufficiently constrained, however, for any one of them to be
compelling.  The additional scalar states that lie below $1 \GeV$ and
which have been interpreted as diquark-antidiquark or meson-meson
states complicate the picture further.

Indeed it is not clear that one should expect a simple
mixing scenario to be able to explain the observed pattern of states. 
The presence of quarks will change the vacuum and
there is no fundamental reason to think that the mass spectrum of QCD
can be approximately described as consisting of the glueballs of the
pure gauge theory, the usual quarkonia and, where these are close in
mass, mixtures of the two. There is, of course, a constellation of
observed phenomena --- the OZI rule, small sea quark effects etc. --- 
that provides some support for the simple mixing scenario. 
Lattice QCD calculations can help to determine whether this is 
indeed so. The fact that one can smoothly decrease the quark mass 
from large values, where the mass spectrum can be easily understood
in terms of glueballs and quarkonia, down to the smaller values 
where the vacuum is very different from that of the
pure gauge theory, provides 
lattice QCD with an advantage over experiment and phenomenology in 
determing the dynamics. (In addition lattice calculations can vary
the volume and the number of colours, which can also be useful in
this context.) For large quark masses the vacuum is close to that
of the pure gauge theory (`gluodynamics') and the quarkonia are heavy, 
so we expect that the states which are most readily visible when
we use purely gluonic 
lattice operators similar to those used in Monte Carlo simulations of 
pure gauge theories, should be just the glueballs of gluodynamics.
If, when we reduce the sea quark mass, the main new effect is
mixing with quarkonia that are nearby in mass, we would expect
that the visible states maintain their overlap onto these same 
operators, except in the range of quark masses where such mixing 
takes place. Moreover we would expect the mass to be constant
except, once again, when mixing occurs. In this situation one
can talk of `glueballs' in QCD. 
For simplicity we will refer to states calculated in the same way 
as in the pure gauge theory as `glueballs' during the remainder 
of this paper, although, of course, whether such language is
justified is precisely one of the things we will be investigating.

Lattice QCD calculations are time consuming and this will limit the
scope of our calculations. The largest lattice we can simulate is
$16^3 32$. To maintain a large enough physical volume means that our
lattice spacing cannot be significantly below $a \simeq 0.1 \fm$. We
use a non-perturbatively ${\cal O}(a)$ improved quark action, so that the
lattice corrections should decrease rapidly, proportionally to $a^2$, as
we decrease $a$. Unfortunately the improvement coefficients are not
known for $\beta < 5.2$ and this means we cannot increase the value of
$a$ significantly beyond $a \simeq 0.1 \fm$.  Thus our calculations
are limited to a narrow range of lattice spacings. This means that we
are unable to say much about the continuum limit of our glueball
masses. It also means that we cannot easily tell whether any
difference that we observe with gluodynamics is due to lattice
corrections or is a genuine physics effect. This is likely to be a
particular problem for the scalar glueball (and perhaps for other
$J^{PC}=0^{++}$ states) which, in gluodynamics at least, possesses
quite large lattice corrections at this value of $a$. This so--called
`scalar dip' is related to a critical point in the plane of
fundamental and adjoint couplings, that is close to the Wilson axis
and where the scalar glueball mass vanishes, e.g.
\cite{creutz81,patel86,heller95a,heller95b,morningstar98,
morningstar99,morningstar99b}. 

Another potential problem concerns the accuracy of our
glueball mass calculations. Glueball (and other flavour
singlet) correlation functions are notorious for their
poor signal to noise ratios, which decrease rapidly
with $t$. In gluodynamics one typically 
needs ${\cal O}(10^5)$ Monte Carlo sweeps in order to extract
anything useful on the lightest glueball masses at
$a \simeq 0.1 \fm$. Whether one can obtain something
meaningful on our modest sized dynamical QCD runs 
is thus far from clear.

One purpose of this paper, then, is to learn more about the
statistical and systematic errors discussed above, so as to
provide useful information for future lattice QCD studies.
We do this by attempting to calculate glueball masses on ensembles 
of lattice QCD fields recently produced by the UKQCD Collaboration
\cite{allton01}.
We find, remarkably, that even on these moderately sized ensembles, 
we can robustly calculate the masses of the lightest glueball
excitations. 

Our second purpose is to learn something about how sea quarks
affect the glueball component of the hadron spectrum, as discussed
above. Here it will be important to find arguments that enable us
to distinguish lattice artifacts from genuine sea quark effects.

The structure of this paper is as follows. In Section~\ref{sec_methods}
we discuss the UKQCD ensembles, and the methods used for estimating
the correlation functions from which we extract string tensions
and glueball masses. (Details are relegated to the Appendix.) 
In Section~\ref{sec_strings} we present our results for the 
flux tubes that wind around a spatial torus (`torelons') 
from which we extract string tensions. We also show that these flux 
tubes become increasingly unstable as the quark mass is reduced
--- an example of `string breaking'. 
In Section~\ref{sec_glueballs} we present our results for the glueball
masses. We compare these to the quenched glueball spectrum at the
same value of $a$ and attempt to disentangle lattice and finite 
volume artifacts from genuine sea quark effects.
Finally we summarise and discuss our results in 
Section~\ref{sec_conclusions}.

For a recent study of the the glueball spectrum in the presence of 
sea quarks which uses an unimproved Wilson action, we refer to
\cite{bali00}.
Recent studies of mixing can be found in
\cite{toussaint00,michael00}.
Results from this study for one value of the quark mass
have been presented in
\cite{allton01},
and in a much more preliminary form in
\cite{irving00}.
\section{Methods}
\label{sec_methods}

\subsection{Ensembles}

In this paper we use six ensembles of SU(3) lattice gauge fields
that have been recently generated by the UKQCD collaboration using a
QCD lattice action with $N_f=2$ degenerate flavours of (dynamical) sea
quarks
\cite{allton01}.
As shown in Table~\ref{tab_ensembles}, these ensembles (labelled
e$_n$) have been produced with two notable features. The first is the
use of an improved action, such that leading order lattice
discretisation effects are expected to depend quadratically, rather
than linearly, on the lattice spacing (just as in gluodynamics). In
addition, the action parameters have been chosen to maintain specific
values of the lattice spacing and quark mass.

The SU(3) gauge fields are governed by the Wilson plaquette action,
with `clover' improved Wilson fermions
\cite{allton01}.
The improvement is non--perturbative, 
with the coefficient $c_{\rm sw}$ chosen to render the
leading order discretisation errors quadratic (rather than linear) in
the lattice spacing, $a$.

The theory has two coupling constants. In pure gluodynamics the gauge
coupling, $\beta$, controls the lattice spacing, with larger values
reducing $a$ as we move towards the critical value at $\beta = \infty$. 
In simulations with dynamical fermions it has the same role for a fixed
fermion coupling, $\kappa$. The latter controls the quark mass, with
$\kappa \to \kappa_c$ from below corresponding to the massless
limit. In dynamical simulations, however, the fermion coupling also
affects the lattice spacing, which will become larger as $\kappa$ is
reduced (and hence $m_q$ increased) at fixed $\beta$.

Ensembles e$_6$, e$_5$ and e$_4$ (in order of decreasing sea quark
mass) have been generated with appropriately decreasing $\beta$ as
$\kappa$ is increased, so the couplings are `matched' to maintain a
constant lattice spacing, as defined by $\hat{r}_0 \equiv r_0/a$
\cite{irving98,garden99},
whilst approaching the chiral limit%
\footnote{We use circumflex accents to denote dimensionless lattice
quantities.}.
This lattice spacing is
`equivalent' to $\beta \simeq 5.93$ in gluodynamics with
a Wilson action. Discretisation and finite volume effects should thus
be of similar sizes on these lattices. Ensembles e$_3$ and e$_2$ are
similarly matched at a slightly finer lattice spacing. In addition,
ensembles e$_3$ and e$_4$ were chosen to have matched chirality (i.e.
to have the same value of $m_\pi / m_\rho$ and hence of the sea quark 
mass) at different lattice spacings with a view to an eventual 
continuum extrapolation.
Finally, we include results for a slightly smaller ensemble
e$_1$, which has the lightest sea quarks at the finest lattice
spacing, `equivalent' to $\beta \simeq 6.00$. For an explicit description 
of the action, see
\cite{allton01}.
An illustration of the parameter space may be found in
\cite{ukqcdsf00}.
The lattices all have a spatial extent $L \gsim 1.5 \fm$, 
which is significantly larger
than many quenched simulations, and this should ensure that any finite
volume corrections are minor. (We will discuss this
further below.)

Four--dimensional lattice gauge theories are scale invariant, and the
dimensionless lattice quantities must be cast in physical units
through the use of a known scale. That is, a specific quantity
measured on the lattice (at a finite lattice spacing) is defined to
agree precisely with its known physical value by multiplying by the
appropriate powers of the lattice spacing. In this work we shall use
(separately) two such scales; the string tension extracted from the
ground state mass of `torelons', single closed periodic flux tubes
that wind around the periodic spatial lattice, and the Sommer
scale,
$r_0$, extracted from potential calculations. We present
here details of the determination of the former, whilst measurements
of the Sommer scale and further details of the UKQCD ensembles are
described in
\cite{allton01}.

Measurements were made on ensembles of $N_{\rm conf} = 280-830$
configurations of size $L^3T = 16^3 32$, separated by ten hybrid Monte
Carlo trajectories.  Correlations in the data were managed through
jack--knife binning of the data, using bin sizes large enough that
neighbouring bin averages may be regarded as uncorrelated. In practice
this amounted to dividing the data into ten bins of length between 280
and 830 trajectories.  Where the Sommer scale has been employed, its
errors are combined in quadrature with those of the mass. For
quantities in units of the string tension, the statistical error
estimate comes from a combined jack--knife bin analysis. It is to the
determination of the string tension that we turn next.

\subsection{Torelons: calculating the string tension}

In a periodic, spatial volume colour flux tubes can close upon
themselves by winding around the spatial torus. If we are in
the low temperature confining phase of gluodynamics such a closed 
loop of flux will be stable. Such a state, with unit net winding
number in one of the spatial directions, is often
referred to as a `torelon'. To a first
approximation the mass of such a state is the spatial extent of the
lattice multiplied by the energy per unit length of the flux tube,
that is the string tension. In the infinite volume limit the
torelons become very massive and decouple from the observed 
spectrum. Including the usual universal string correction
\cite{forcrand85},
we expect the torelon mass to vary with the lattice size as
\be
\hat{m}_\tau = \hat{K}L - \frac{\pi(D-2)}{6L},
\label{eqn_sigma}
\ee
where $\hat{K}$ is the lattice string tension and $D$ is the
dimensionality of space-time.  We estimate the mass of this state by
measuring correlations of spatial Polyakov loop operators
\be
\tau_\mu(n) = \Tr \prod_{k=1}^L U_\mu(n+k\hat{\mu}).
\label{eqn_poly}
\ee
for $\mu = 1...3$. These are summed in various momentum combinations
up to $P \cdot P = 2$ in units of the lowest momentum, $\frac{2
\pi}{L}$. We improve the overlap of the operator onto the ground state
by iteratively smearing and blocking the gauge fields as described in
Appendix~\ref{sec_improvement}. Improvement levels $i=5,6$ were found
to best saturate the ground state overlap (but we could include all
levels in a $7\times7$ correlation matrix without loss of robustness).

This simple description becomes more complicated in QCD, where 
colour flux tubes can break. If the decay width is small, the
flux tube will still be a well-defined, albeit unstable, state.
In that case we will be able to proceed with the above analysis
with some modifications, as described in detail later on.

\subsection{Calculating glueball masses}

For the purposes of this paper we shall concentrate on the lightest
glueball states only, as we possess neither large enough ensembles, 
nor small enough (timelike) lattice spacings for a comprehensive 
calculation of the spectrum. As a rough estimate, and by analogy 
with gluodynamic calculations at the same value of $a$, we could, 
at best, hope to make accurate determinations of the scalar mass 
(the $A_1^{++}$ which becomes $J^{PC}=0^{++}$ in the continuum
limit). We may also be able to obtain some estimates of the tensor
($T_2^{++}$), and perhaps also the pseudoscalar ($A_1^{-+}$), the 
vector ($T_1^{+-}$) and the first excited state of the scalar.
We will, in fact, find that while we can estimate the scalar 
and tensor ground state masses, we can do no more than put upper 
bounds on the higher states.

To calculate the mass of a glueball excitation with given quantum
numbers, it is necessary to measure correlation functions of operators
with the same symmetry properties (referred to as being `in the same
channel') 
\cite{michael90},
and with improved overlap onto the low--lying states of that channel.
The operators at each site are summed over all spatial sites in a
timeslice to create momentum eigenstates for 3--momenta up to $P \cdot
P = 2$.

Gauge links are improved as in Appendix~\ref{sec_improvement}, and we
extract masses either from zero--momentum effective masses or, in the
case of the light scalar, by variational analysis of a reduced, $8
\times 8$ correlation matrix utilising momenta $P \cdot P = 0,1$. This
is explained in Appendix~\ref{sec_extraction}, where details of fits
to individual states are given.

\section{Torelons and the string tension}
\label{sec_strings}

In the confining phase of gluodynamics the Polyakov loop operator shown
in Eqn.~\ref{eqn_poly} has zero overlap onto a contractible loop
of the kind one uses as a trial glueball wave-functional. This
implies that a winding flux tube has zero probability to turn
into a glueball. Physically it is a consequence of the fact that
the flux tube cannot break. Mathematically it is a consequence
of the following symmetry argument. Suppose the Polyakov loop winds 
once around the $x$-torus. Choose some value of $x$, say $x_0$, 
and multiply all the link matrices in the direction $x$ at
$x=x_0$ by $z_i$, a non-trivial element of the centre of the
group, $Z(3)$, i.e.
\be
U_x(x_0,y,z,t) \to z_i U_x(x_0,y,z,t) \quad \forall \, \, y,z,t .
\label{eqn_centre}
\ee
Let the original and transformed fields be labelled 
$\{U_l\}$ and $\{U^z_l\}$ respectively. Since the Haar integration
measure $dU_l$ and the plaquette action are invariant under
the transformation in Eqn.~\ref{eqn_centre}, the fields
$\{U_l\}$ and $\{U^z_l\}$ appear with exactly the same weight in 
the path integral. Now in a contractible loop, such as the 
plaquette, any factor of $z_i$ from a forward going link is 
necessarily matched by a corresponding factor of $z_i^\dagger$ from 
a backward going link -- so the value of the loop is unchanged.
By contrast, $\tau_x$  will acquire just a single factor of $z_i$.
The argument applies to any element of the centre and therefore
the vacuum expectation value (VEV) of the Polyakov loop will satisfy
\be
\langle \tau_x \rangle 
= 
\sum_{z\in Z_3} z \langle \tau_x \rangle = 0 .
\label{eqn_vacpoly}
\ee
A similar calculation shows that the correlation function of
a Polyakov loop with a glueball operator is also zero.
This argument only breaks down if the symmetry is spontaneously
broken, so that the the path integral becomes restricted
to a subset of fields, and the vacuum ceases to be symmetric.
This occurs, of course, at high temperature where confinement
is lost, if we consider Polyakov loops, $\tau_t$, that wind 
around the Euclidean time torus. By the same token it will
occur for a Polyakov loop that winds around the $x$-torus,
as here, if the size of this torus is reduced below a
critical value. 

All the above remains true for blocked or smeared operators. 
For instance, on quenched $16^4$ lattices 
at $\beta=5.93$ (where $\hat{r}_0$ is approximately
the same as on our dynamical ensembles e$_4$, e$_5$ and e$_6$) we
find $\langle \tau \rangle = (7 \pm 13) \times 10^{-6}$. Upon
blocking the value remains consistent with zero, being $(7 \pm 241)
\times 10^{-6}$ after four blocking steps.

In QCD (always at low $T$) the symmetry in Eqn.~\ref{eqn_centre}
is broken explicitly by the fermionic piece of the action, because
the latter is linear in the link matrices. We can estimate the
importance of this effect as follows. Suppose we integrate out
the fermionic fields (as indeed one does in practice). The
result is a partition function that is just as in the pure gauge 
theory except that the integrand acquires an extra  factor,
the determinant of the Dirac operator $\det (D(U)+m_q)$, for
each fermion flavour. The $\det (D(U)+m_q)$ can be expanded as
a sum over closed loops and we can divide this sum into
two pieces: $\det_{\rm C}$ that consists of the (contractible) closed 
loops that transform trivially under Eqn.~\ref{eqn_centre},
and  $\det_{\rm NC}$  that consists of the (non-contractible) closed 
loops that transform non-trivially under Eqn.~\ref{eqn_centre}.
The latter contains products of closed loops, one of which
must wind once around the $x$-torus. Thus, specialising
for simplicity to the case of one flavour,
\begin{eqnarray}
\langle \tau_x \rangle 
& = &
{1\over Z} \int \prod_l dU_l \, \tau_x \,
\left( {\rm det}_{\rm C} \left(D(U)+m_q \right) + 
{\rm det}_{\rm NC} \left( D(U)+m_q \right) \right) e^{-S_G} \nonumber \\
& = &
{1\over Z} \int \prod_l dU_l \, \tau_x \, 
{\rm det}_{\rm NC} \left( D(U)+m_q \right) e^{-S_G} ,
\label{eqn_vacpolyqcd}
\end{eqnarray}
where $S_G$ is the usual plaquette action.
If the quark mass is large, the winding loop contributes
$\propto (1/m_q)^L$ and loops are rare so that we can
factorise $\det_{\rm NC} \propto (1/m_q)^L \times \det_C$
and so we find
\bea
\langle \tau_x \rangle 
& = &
{1\over Z} \int \prod_l dU_l \tau_x 
{\rm det}_{\rm NC} \left( D(U)+m_q \right) e^{-S_G}
\nonumber \\
& \propto &
- \left( {1\over{m_q}} \right)^L .
\label{eqn_vacpolyqcd_largem}
\eea
The minus sign is because we have a closed fermion loop.
(We have periodic spatial boundary conditions on the fermions.)
All this is true not only for simple loops but also for
smeared Polyakov loops. The argument extends trivially to
any spatial direction and to any number of flavours.

Eqn.~\ref{eqn_vacpolyqcd_largem} embodies our expectation that 
as the sea quark mass becomes large, the quenched limit is approached
and the VEV goes to zero. As the sea quark becomes lighter
the VEV will, at first, rapidly increase, but at some point
the mass in the winding loop will approach 
a value that reflects the finite constituent quark mass in the 
theory, and thereafter the VEV will no longer vary so rapidly.

A simple Polyakov loop can be regarded as the propagator of
a heavy fundamental source and we may thus use
$\langle \tau_x \rangle \propto \exp -(\hat{F}L)$ to 
calculate a `spatial' quark free energy. Since a non-zero
value of $\langle \tau_x \rangle$ requires a quark to
propagate around the lattice, we may also use its value
to determine the mass of the corresponding heavy-light system.
We are here primarily interested in the flux tube and
this couples to suitably smeared Polyakov loops for which
such interpretations are not useful, however, and we thus do not develop 
them any further here.

The above discussion justifies the usual intuitive picture of 
how $ \tau_x$ acquires a non-zero vacuum expectation value.
One thinks of a $q\bar{q}$ pair popping out of the vacuum
somewhere along the flux tube, separating and thus breaking the
flux tube, and finally annihilating at the boundary, at which
point the flux tube has disappeared. The same process could,
of course, leave us with a closed, contractible flux tube, which 
is a potential wave-functional for a glueball. So we might expect 
the amplitude for glueball-flux tube mixing to be of the same
order as the vacuum expectation value. If the $q\bar{q}$
do not annihilate round the back of the torus then we are
left with a potential quarkonium state. So, to the extent that
the annihilation is suppressed, we can expect the amplitude
for mixing with quarkonia to be larger.  

\begin{figure}[p]
\begin{center}
\leavevmode
\epsfysize=300pt
\epsffile{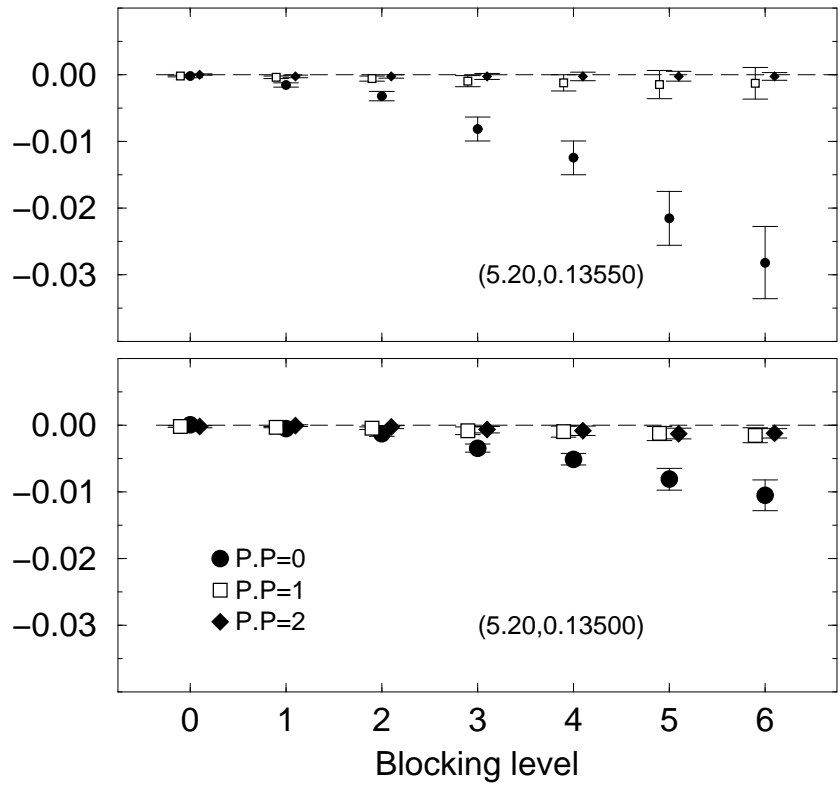}
\end{center}
\caption[]{\label{fig_poly_vevs} {\it Vacuum expectation values for
  Polyakov loops at various blocking levels on the e$_2$ and e$_4$
  ensembles. Plotting point radii are proportional to $\hat{r}_0
  \hat{m}_\pi$.}}
\end{figure}

In Fig.~\ref{fig_poly_vevs} we plot the vacuum expectation value 
of the Polyakov loop for the representative ensembles e$_2$ and 
e$_4$. We observe a small vacuum expectation value that is 
nonetheless statistically significant. Of course this assumes
that we have not underestimated our statistical errors.
To demonstrate this we also plot the vacuum expectation values
of Polyakov loop operators with $P \cdot P=1,2$. From
momentum conservation we expect the VEVs of these
operators to be zero, and this we see is satisfied within less than
two standard deviations in all cases. This shows that our
statistical errors estimates are indeed reliable. We can thus
claim to have obtained clear evidence of flux tube breaking by 
dynamical quark pair production, analogous to the string breaking 
that one tries to see in the static quark potential, as calculated
using Wilson loops.

We note that the VEV of the simple Polyakov loop is consistent
with zero and that it only becomes significant as the
blocking level is increased. Since the overlap onto the
lightest flux tube also increases rapidly with increasing 
blocking level, we interpret this as telling us that 
what we are seeing is the breaking of this lightest
flux tube. We also note that the VEV is larger for the
ensemble corresponding to the lighter value of $m_q$.
This is what we would intuitively expect for string
breaking and is consistent with eqn(\ref{eqn_vacpolyqcd_largem}).

\begin{figure}[p]
\begin{center}
\leavevmode
\epsfysize=300pt
\epsffile{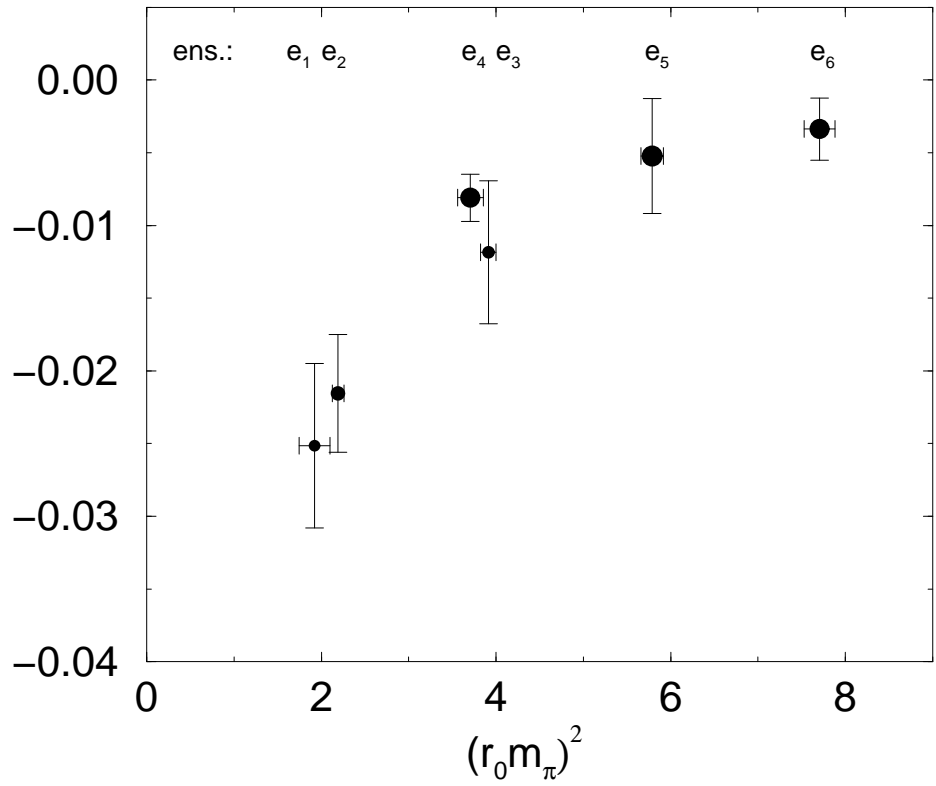}
\end{center}
\caption[]{\label{fig_poly_vevs_mq} {\it Vacuum expectation values for
  Polyakov loops at the optimal blocking level (5 in each case) versus
  the squared pion mass. Plotting point radii are proportional to
  $\hat{r}_{0}^{-1}$.}}
\end{figure}

This last point is explored in more detail in 
Fig.~\ref{fig_poly_vevs_mq}. Here we plot the VEV at the $i=5$ 
blocking level versus the quark mass (using the fact that
$m_q \propto m^2_{\pi}$). We choose the  $i=5$  blocking level
since this has the largest overlap onto the lightest flux tube. 
We clearly observe an increase in the VEV as the quark mass
is decreased. Indeed the quark masses at which the effect turns
on rapidly is broadly consistent with the mass where the
suppression of the topological susceptibility becomes significant
\cite{hart99,allton01}.
This is corroborative evidence for a significant change in the
properties of the QCD vacuum generated by light sea quarks, a point
to which we shall return later.

As we have already remarked, the string breaking we see,
while significant, is weak. We reinforce this observation
by listing in Table~\ref{tab_mixing} the overlap of torelons 
in orthogonal spatial directions
\be
M_{\mu \nu}(t) = \frac{\langle \tau_\mu(t) \tau_\nu(0) \rangle}
{\sqrt{\langle \tau_\mu(t) \tau_\mu(0) \rangle
\langle \tau_\nu(t) \tau_\nu(0) \rangle}}
\ee
which we see is consistent with zero. The fact that these
overlaps are zero also means that nothing is to be gained by
producing linear combinations that transform according to
irreducible representations of the rotation group. 

The above tells us that while the lightest flux tube is visibly 
unstable in the presence of sea quarks, and increasingly so as the
quark mass decreases, its decay width, for our range of quark masses,
is small. We may therefore use Eqn.~\ref{eqn_sigma} to interpret the 
torelon mass in terms of the string tension, just as we would
in the quenched theory. Doing so for $P \cdot P=0,1$ (using a
variational analysis of the connected correlation matrix, as
explained in Appendix~\ref{sec_extraction}) we obtain the torelon 
masses and string tensions listed in Table~\ref{tab_toron_fits}.
We remark that these string tensions are in good agreement with 
those that have been obtained from Wilson loop measurements in
\cite{allton01}.
\section{Glueballs}
\label{sec_glueballs}

The scalar glueball mass is well determined in our calculations
and we list our results for it in Table~\ref{tab_scalar_fits}.
The tensor glueball is also quite well constrained, as we see in
Table~\ref{tab_tensor_fits}. These scalar and tensor masses are 
expressed in units of the Sommer scale and the string tension in
Table~\ref{tab_scalar_masses} and Table~\ref{tab_tensor_masses}
respectively. The masses of other states
are listed in Tables~\ref{tab_ex_scalar_fits},~\ref{tab_pseudo_fits}
and \ref{tab_vector_fits}. The errors on the masses are clearly
large, and in most cases the evidence for a plateau in the effective
mass for increasing $t$ is unsatisfactory. In the interest of
completeness we give our estimates for the first excited scalar, the
pseudoscalar and vector glueballs, in units of $r_0$ and $\sqrt K$, 
in Table~\ref{tab_other_masses}, but
we stress that these should be regarded either as crude estimates
or as upper bounds.

We shall begin by comparing our results with quenched calculations
at the same value of $a$, so as to establish what, if any, are
the effects of the sea quarks. We shall then ask whether the
observed effects might not be due to finite volume effects, in
particular to those which are only present once we have
sea quarks. We shall see that such effects can be excluded.
We then discuss the implications of what we find for glueball
masses in QCD.

\subsection{Quenched comparison}
\label{subsec_interpolation}

To identify what effect the sea quarks have on our calculated glueball
masses, we will now compare our results to those obtained 
in the pure gauge theory. Since our lattice spacing is quite coarse,
we can expect lattice spacing corrections to be significant, and
so we will make the comparison at the same value of $a$.

Since we are comparing two different theories, there is, strictly 
speaking, no such thing as the `same' value of $a$. We shall, 
however, follow usual practice and use the Sommer scale, $r_0$, to 
calibrate the lattice spacing. That is to say, we calculate $r_0/a$
in the theory with and without quarks and say that $a$ is the
same when $r_0/a$ is the same. That this is sensible is based
on arguments that $r_0$ should be insensitive to the presence of
sea quarks.

Although, in principle, $r_0$ provides a useful scale, the fact that
it is extracted from the static potential at intermediate 
distances ($\sim 0.5 fm$) means that different ways of calculating
it will differ when $a$ is coarse. To avoid this (probably minor)
systematic error we shall compare with the quenched calculation
at
$\beta=5.93$ where $r_0/a = 4.714(13)$
\cite{allton01}
using exactly the same procedures as for the QCD calculations
\cite{allton01}.
As we see from Table~\ref{tab_ensembles}, this lattice spacing
is the same (within errors) as the one for the ensembles
e$_4$, e$_5$ and e$_6$. The lightest glueball masses and the
string tension have been calculated on a $16^4$ lattice at
$\beta=5.93$ in 
\cite{lucini01}.
The value of the string tension obtained there is
$\surd{\hat{K}}=0.2430 \, (23)$ which agrees well with the values 
we obtain on the e$_4$, e$_5$ and e$_6$ emsembles, as quoted 
in Table~\ref{tab_toron_fits}. The values of $a$ are slightly
smaller on the other ensembles, with the smallest being on the e$_1$ 
ensemble. If we scale up the quenched scale by the same amount
(using the string tension, which should be a reliable procedure
over this small range of scales) we conclude that the equivalent
quenched $a$ to the e$_1$ ensemble is at $\beta \simeq 6.0$. 
Quenched calculations at this value of $\beta$, on a $16^4$ lattice, 
have been performed and can be found in
\cite{lucini01}.
In the quenched theory, at $\beta=5.93$, the lightest scalar and
tensor glueball masses are $\hat{r}_0 \hat{m}_G = 3.68 \,(8)$ and 
$\hat{r}_0 \hat{m}_G = 5.82 \, (14)$ respectively. At $\beta=6.0$
the corresponding values are  $\hat{r}_0 \hat{m}_G = 3.83 \,(8)$ 
and $\hat{r}_0 \hat{m}_G = 5.90 \, (14)$. (The $\beta=6.0$ value of
$\hat{r}_0$ has been estimated by scaling the $\beta=5.93$ value
by the ratio of the calculated string tensions.) That is to say,
the tensor mass is unchanged, while the scalar increases
slightly as $\beta$ increases. We shall thus take the the
$\beta=5.93$ masses to represent the equivalent quenched values 
for all our six QCD ensembles.

In Fig.~\ref{fig_compare_masses} we plot our calculated values 
of the lightest scalar and tensor glueball masses as a
function of the square of the pion mass, which should be 
approximately proportional to the sea quark mass. We also plot 
the equivalent quenched masses (obtained at $\beta=5.93$
\cite{lucini01})
and, in addition, their continuum extrapolation
\cite{morningstar99}.
We observe that the scalar mass is systematically lighter than
its quenched value. If we average over the six ensembles 
we obtain $0.85 \, (3)$ for the ratio of the dynamical to quenched 
scalar masses. For the tensor the corresponding ratio is 
$0.98 \, (4)$ and here we conclude that there is no sea quark effect
visible within the (not so small) errors. 

Our mass estimates
for the lightest pseudoscalar, the first excited scalar, and the
vector have very large errors making any comparison not very 
significant. Nonetheless we note that the values listed in
Table~\ref{tab_other_masses} are broadly consistent with the
continuum quenched values obtained in
\cite{morningstar99}.
\subsection{Finite volume effects}
\label{subsec_volume}

As we have just seen, our scalar glueball is significantly
lighter than it is in gluodynamics at the same value of $a$.
Before dwelling upon the implications of this observation
it is important to establish that this effect is not a
mere finite volume effect.
 
In gluodynamics, sizeable finite volume corrections first
arise when the spatial volume decreases to the point where
the mass of a (conjugate) pair of winding flux tubes
becomes less than the value of the glueball mass on large
volumes. In the first part of this Section we shall 
demonstrate that this effect is not significant for the scalar 
glueball on the $L \simeq 3 \hat{r}_0$ lattices considered here. 

In the presence of sea quarks the flux tube can break and this 
makes it possible
for the glueball to mix with a single winding flux tube. This 
could, potentially, introduce finite volume corrections on
larger volumes. We shall examine this mechanism and
show that it is negligible in the present calculation.

\subsubsection{Mixing with torelon pairs}
\label{subsubsec_torelon}

In the pure gauge theory the onset, as we decrease the volume, 
of sizeable finite volume effects,
comes from mixing between glueball states and `torelon pairs'. These
are combinations of two flux tubes with net zero winding number. An
example of a scalar torelon pair operator is 
$T(t) = T_1(t) + T_2(t) +T_3(t)$, where
\be
T_\mu(t;P=0) = \tau_\mu(t;P=0) \tau_\mu^\dagger(t;P=0) .
\ee
(No sum over indices is implied and vacuum subtraction is understood.) 
It is clear that in a large volume the mass of this state should 
be close to twice the torelon energy, and proportional to the
spatial extent of the lattice. When the volume is small enough, the
mass of such a torelon pair becomes first comparable to and then
less than the (large volume) glueball mass. It then
effectively becomes the lightest glueball state (for an analysis in
$d=2+1$, see
\cite{teper98b}).
On our lattices, however, a single torelon is already slightly heavier 
than the scalar glueball, so the torelon pair is likely to be about
twice as heavy. Thus any mixing with the scalar will be heavily
suppressed and the mass itself would be expected to be too heavy to
be visible in our calculations. Given the high mass, it
seems unlikely that any of the glueball states that we consider will
suffer finite volume effects from such torelon pairs.

\subsubsection{Mixing with single torelons}
\label{subsubsec_toron}

Mixing between the scalar glueball and the torelon pair is suppressed,
in our calculation,
by the large mass difference between the states.  As we have seen, the
presence of sea quarks in the simulation breaks the centre symmetry,
albeit weakly at the quark masses accessible to us in this study. It
is now possible for mixing to occur between single torelons and glueballs
\cite{kripfganz89}. 
The torelons are, for a fixed lattice size, much lighter than the
torelon pairs and this can lead to enhanced mixing and finite volume
effects in dynamical simulations. A measure of the strength of the
mixing is the normalised cross correlation of the `best' scalar with
the `best' winding $A_1^{++}$ operator:
\be
M_{\tau \phi}(t) = 
\frac{\langle \tau(t) \phi(0) \rangle}
{\sqrt{\langle \tau(t) \tau(0) \rangle \langle \phi(t) \phi(0) \rangle}}.
\label{eqn_torglue}
\ee
The measurements for $t=0$ are tabulated in Table~\ref{tab_mixing},
and are clearly very small, and indeed consistent with zero. 
(As remarked earlier, the fact that orthogonal torelons have
zero overlap, within errors, means that nothing was
gained by using in Eqn.~\ref{eqn_torglue} the scalar sum
of the three orthogonal torelons.) Thus single torelons will
not lead to any significant finite size corrections to our
scalar glueball masses.

\subsection{Glueballs with sea quarks}
\label{subsec_gluequark}

Now that we have established that the observed suppression of the 
scalar glueball mass is not a finite volume effect, we can turn
to a detailed discussion of its dynamical implications.

\begin{figure}[p]
\begin{center}
\leavevmode
\epsfysize=300pt
\epsffile{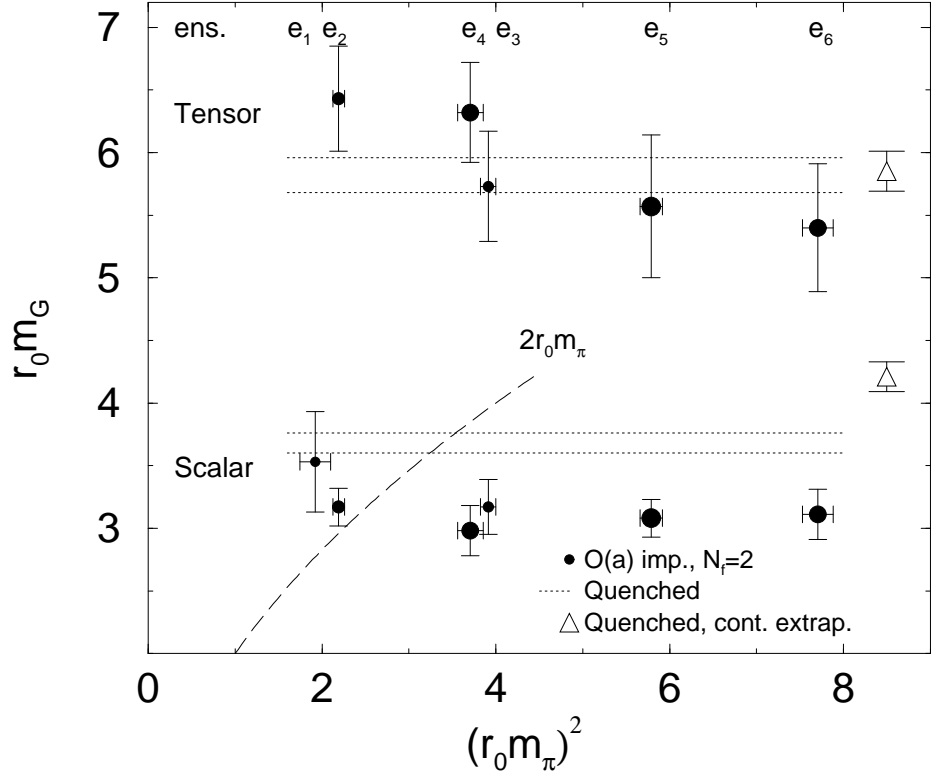}
\end{center}
\caption[]{\label{fig_compare_masses} {\it Comparing dynamical and
  quenched glueball masses. The dashed line is the threshold for decay
  to $\pi \pi$. Dynamical data plotting point radii are proportional
  to $\hat{r}_{0}^{-1}$. The horizontal bands are the quenched masses
  at an equivalent lattice spacing, $\beta = 5.93$
  \cite{lucini01,allton01}. The quenched continuum limits are from
  Ref.~\cite{morningstar99}. N.B. $\hat{r}_0 \hat{m}_\pi$ is not
  applicable to the quenched data.}}
\end{figure}

The first and rather striking feature of the scalar mass 
suppression, as displayed in Fig.~\ref{fig_compare_masses}, is that 
it appears to be independent of the quark mass. This is puzzling
because our range of quark masses goes well above twice the 
strange quark mass where any quarkonia are expected to be much 
heavier than the scalar glueball and so any mixing effects
should be very small. As for the vacuum, we know from our calculations
of the topological susceptibility that sea quarks only become
important about mid-way along the plot, where $\hat{r}_0 \hat{m}_\pi \leq 2$
\cite{hart99}.
(We also see this in Fig.~\ref{fig_poly_vevs_mq} for the torelon VEV.)
Thus it would seem difficult to argue that
this mass suppression is a dynamical effect due to `light' quarks.

The only sizeable effects that one would expect to depend weakly on
the quark mass are lattice artifacts. One can indeed make
a plausible argument for this in the present case. As we
have previously remarked, at this value of $a$ the scalar
mass is already suppressed in gluodynamics by the presence
of a nearby critical point in the space of fundamental
and adjoint couplings. If the effect of our fermionic
action is effectively to add a positive adjoint piece to the
pure gauge action, then this could strongly enhance the
suppression, just as we observe. In fact one can show 
\cite{hart_prog}
that the clover term, inserted to provide the ${\cal O}(a)$ improvement,
has precisely this effect. We believe this to be the most likely
explanation for the supression of the scalar glueball mass.

\begin{figure}[p]
\begin{center}
\leavevmode
\epsfysize=300pt
\epsffile{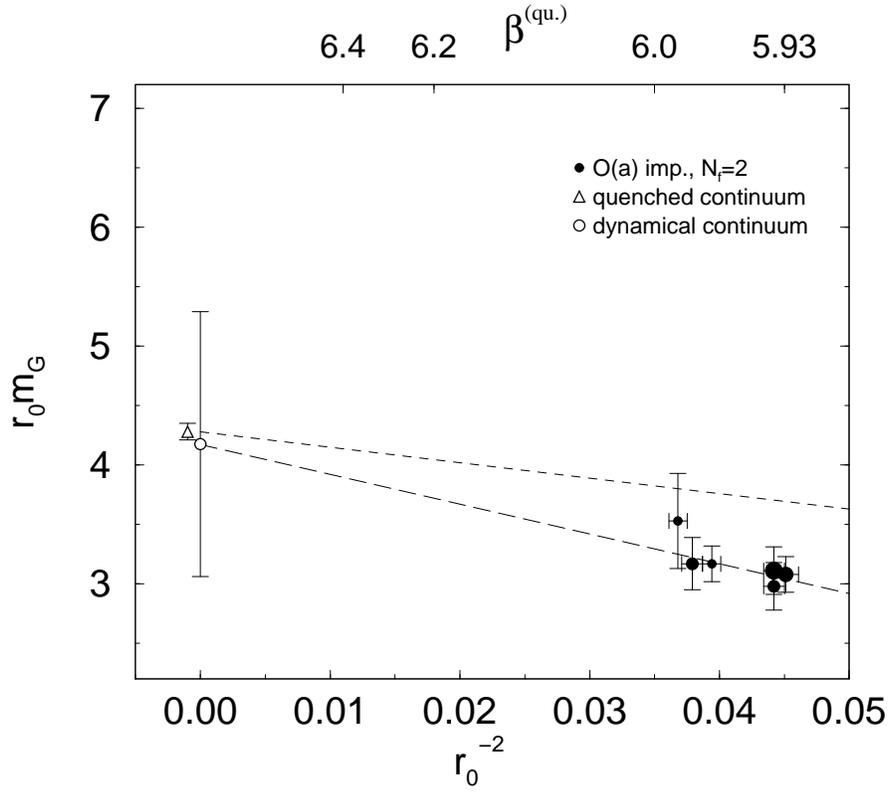}
\end{center}
\caption[]{\label{fig_cont_masses} {\it Discretisation effects on
  scalar glueball mass. Dynamical data plotting point radii are
  proportional to $\hat{r}_0 \hat{m}_\pi$. Also shown is the quenched
  extrapolation, and couplings, $\beta^{\rm (qu.)}$ yielding these
  values of $\hat{r}_0$.}}
\end{figure}

If this suppression is a lattice artifact, then it should
disappear as we take the continuum limit. Our range of 
$a$ is very limited of course, but our lever arm is enhanced by
the fact that the leading corrections, with our action,
are ${\cal O}(a^2)$. Using the fact that the glueball mass appears
not to depend on the quark mass, we plot all our scalar 
glueball masses against $a^2$, in Fig.~\ref{fig_cont_masses}.
We might, very optimistically, claim to see some 
hint here that the suppression decreases with decreasing $a$. 

Let us assume, as argued above, that the suppression at large quark
masses is a lattice artifact. Such a lattice artifact will be largely
independent of the quark mass and so, if we correct the scalar
glueball mass for it, we will obtain a scalar mass that does not vary
significantly with $m_q$.  In particular there will be no change at
$\hat{r}_0 \hat{m}_\pi \sim 2$ where we have observed the onset of
large light-quark effects in quantities such as the topological
susceptibility. Thus we conclude that the scalar glueball is not
sensitive to the increasing presence of quark loops in the vacuum as
these become lighter.  That is to say, the glueball remains a
glueball.

Finally, what of mixing with quarkonia? We do not have much
information on the masses of flavour singlet quarkonia 
for our ensembles. What we do have is the mass
of two pions (in an s-wave), however. The threshold mass is just
$2 m_\pi$ and this is plotted in Fig.~\ref{fig_compare_masses}.
We see that this state is almost degenerate with the 
scalar glueball from the e$_2$ ensemble. The fact that
the mass shows no shift suggests little coupling between 
these states. This would be consistent with a weak
mixing picture of the kind that has usually been assumed
in phenomenological analyses, such as
\cite{lee99,close01}.

All these observations are supported by the fact that 
our best glueball operator does not vary significantly
with $m_q$ and neither does the overlap of the lightest
scalar state on this operator. This suggests that the nature 
of the glueball wave-functional is insensitive to the quark mass.

\section{Conclusions}
\label{sec_conclusions}

One aspect of our analysis that was a (pleasant) surprise
was the relatively small statistical errors that we 
achieved even with our lightest quark masses. As an example
we show in Table~\ref{tab_statistics} how the scalar
glueball effective mass, and its statistical error,
compared with what one obtains in the quenched calculation.
(The latter on a $16^4$ rather than a $16^3 32$ lattice.)
We recall that the QCD calculation makes measurements every 
tenth trajectory in a run that totals 8300 HMC trajectories,
while the quenched
calculation  makes measurements every fifth of
$10^5$ Monte Carlo sweeps. The difference in the errors
is surprisingly modest and makes us optimistic about 
future QCD calculations of flavour singlet hadron masses.

We calculated the masses of what we hoped would turn out
to be flux tubes winding around the spatial torus. We
found that such flux tubes are indeed unstable, as one
would expect in QCD, but this effect is weak enough
for us to obtain accurate values of the string tension,
which  agree with the values obtained in the pure gauge theory
with the same value of the lattice spacing. The instability
of the flux tube is signalled by a non-zero vacuum expectation
value of the best flux tube wave-functional, whose onset
becomes marked around the strange quark mass. This may
be viewed as a direct example of the expected `string-breaking'.

In our glueball mass calculations we found that we could
obtain reliable results for the lightest scalar and tensor
glueballs. The tensor mass shows no dependence on the
quark mass and its value, $\hat{r}_0 \hat{m}_G \simeq 6.0\, (5)$,
is consistent with what one finds in the pure gauge theory at the 
same value of $a$, i.e. $\hat{r}_0 \hat{m}_G = 5.82 \, (14)$.
The scalar mass also shows no quark mass dependence
but its value, $\hat{r}_0 \hat{m}_G \simeq 3.0 \, (2)$, is 
significantly suppressed compared to the pure gauge theory
value of $\hat{r}_0 \hat{m}_G = 3.68 \, (8)$. We argued
that the quark mass independence of this suppression, even
at quark masses as large as $m_q \sim 2m_s$, suggests that
this is a lattice artifact rather than an effect due to
`light' quarks. This is made plausible by the observation
\cite{hart_prog},
that the large clover term in our improved action, pushes the
effective gauge action closer to the nearby critical point in the
space of fundamental and adjoint couplings, and thus strongly enhances
the ${\cal O}(a^2)$ lattice corrections.  A similarly low scalar
glueball mass has also been seen at coarser lattice spacings
for the clover action
\cite{michael00}.

Assuming this explanation to be correct leaves no room
for any significant variation of the scalar mass from its
quenched value as the sea quarks become important in the 
vacuum. (We infer from the behaviour of the topological
susceptibility and the flux tube vacuum expectation value that 
this occurs at about $\hat{r}_0 \hat{m}_\pi \sim 2$.) This suggests that
the nature of the glueball suffers no major change as
light sea quarks `switch on'. It also leaves no room
for any significant variation at the lightest quark masses
where the two pion s-wave state becomes slightly
lighter than the scalar glueball and where any strong
coupling would presumably lead to a visible mass shift. This
suggests the coupling is weak. All this fits in with the
usual na\"{\i}ve picture of the fate of glueballs in QCD:
they are largely unaffected by the light quarks in the 
vacuum and mix weakly with quarkonia, so that only
nearby quarkonia are important.

It is, of course, the case that we have read a great deal into a 
relatively limited set of calculations and so the above analysis should
be regarded as setting a framework for future discussion, rather
than being in any way definitive.

\section*{Acknowledgments}

This work was performed under the British PPARC research and travel
grants to the UKQCD Collaboration. We thank D. Hepburn and D. Pleiter
for preliminary estimates of pseudoscalar meson
masses. A.H. acknowledges the support of PPARC grants for postdoctoral
support.

\newpage

\appendix
\section{Computational procedures}
\label{sec_appendix}

In this appendix we give details of the procedures used to obtain our
mass estimates.

\subsection{Improvement of operators}
\label{sec_improvement}

To reduce statistical errors on mass estimates, it is necessary to
ensure that operators have a good overlap onto the ground state
excitation with the specified quantum numbers. This was achieved by
a combination of smearing and blocking of the spatial links only.

Smearing
\cite{albanese87}
is computationally cheap, and does not reduce the size of the
effective lattice. The smeared link is a combination of the original
link and the (spatial) staples surrounding it.
\begin{eqnarray}
U_\mu(n) \to U^\prime_\mu(n) & = &
c_{\rm sm} U_\mu(n) 
\nonumber \\
& + & \sum_{\nu=1..3 \not=\mu} \left( 
U_\nu(n) U_\mu(n+ \hat{\nu}) U^\dagger_\nu(n+ \hat{\mu}) \right. 
\nonumber \\
& + & \left.
U^\dagger_\nu(n- \hat{\nu}) U_\mu(n- \hat{\nu})
U_\mu(n+ \hat{\mu}- \hat{\nu})
\right)
\end{eqnarray}
where $c_{\rm sm}$ is a free parameter. The smeared link is then
projected back onto the gauge group.

Whilst smearing alone will couple well to the spherically symmetric
scalar glueball ground state, for the tensor and torelon we find it
important to include blocking steps
\cite{teper87}
in the improvement schedule, where we increase the effective lattice
spacing by a factor of two, and replace each spatial link by
\begin{eqnarray}
U_\mu(n) \to U^\prime_\mu(n) & = & 
c_{\rm bl} U_\mu(n) U_\mu(n+ \hat{\mu}) 
\nonumber \\
& + & \sum_{\nu=1..3 \not=\mu} \left( 
U_\nu(n) U_\mu(n+ \hat{\nu}) U_\mu(n+ \hat{\nu}+ \hat{\mu}) 
U^\dagger_\nu(n+ 2\hat{\mu}) \right. 
\nonumber \\
& + & \left.
U^\dagger_\nu(n- \hat{\nu}) U_\mu(n- \hat{\nu}) 
U_\mu(n+ \hat{\mu}- \hat{\nu})
U_\mu(n+ 2\hat{\mu}- \hat{\nu})
\right)
\end{eqnarray}
where again $c_{\rm bl}$ is a free parameter and the blocked link is
projected back onto the gauge group using a link-wise cooling
procedure. Over-applying this can lead to the creation of supposedly
local objects that in fact have a winding round the lattice, which we
avoid.

We choose $c_{\rm sm} = 1.0$, $c_{\rm bl} = 1.0$ and up to six levels
of iterated improvement are applied to the links, alternately smearing
and blocking (beginning with the former). We label these sequentially
$i=1..6$, with $i=0$ being the original links.

\subsection{Extraction of mass estimates}
\label{sec_extraction}

Using the suite of operators, we form for each momentum channel a
matrix of (vacuum subtracted) correlators $C_{jk}(t) \equiv \langle
\phi_j(t)^\dagger \phi_k(0) \rangle$, normalised such that the
diagonal elements are unity at $t = 0$.  From the diagonal elements we
derive the effective masses, all tending to the ground state for
large $t$,
\be
\hat{m}^{\rm eff}_k(t) = - \ln \frac{C_{kk}(t+1)}{C_{kk}(t)}
\ee
and the overlap of the state onto the ground state
\be
b_k = \frac{C_{kk}(1)^2}{C_{kk}(2)} \le 1
\ee
Most of the states we attempt to study are massive enough that the
correlation functions decay rapidly into the noise. For such states we
derive the mass estimate from the plateau in $t$ of the operator that
best couples to the ground state (i.e. which maximises $b_k$ and
minimises $\hat{m}^{\rm eff}_k(t=1;P=0)$).

The lightest states permit a refined estimate of the low lying masses
via a variational analysis on the correlation matrix
\cite{luescher90,michael00}
(see
\cite{hart00b}
for precise details of our method). For the torelon we have a
$7\times7$, whilst for the scalar glueball we create a reduced
$8\times8$ matrix, using the four basic operators at optimal
improvement levels $i=3,4$. We have also checked that reducing the
basis even further, to the best two states or to simply the diagonal
elements of the correlation matrix gives consistent ground state
determinations.

The generalised eigenvalue problem is solved at each momentum value
for the $f=1...N$ right eigenstates
\be
\sum_{k,m} C^{\dagger}_{jk}(t_0) C_{km}(t) v^{(f)}_m(t,t_0) =
\lambda^{(f)}(t,t_0) v^{(f)}_j(t,t_0).
\ee
where $t_0=0$ in all cases here. In common with most implementations,
we calculate the eigenvectors with normalisation
\be
\sum_{k,m} v^{(f)}_k(t,t_0) C_{km}(t_0) v^{(f)}_m(t,t_0) = 1
\ee
at $t=t_1$ and freeze them at all subsequent times. This improves the
stability of the effective mass estimates. The results presented here
are for $t_1 = 1$ (in lattice spacing units), but we have checked for
$t_1=2,3$ that any biases introduced are within the statistical
errors.  We can approximate the eigenstates of the
Hamiltonian by $\Phi^{f}(t) = \sum_k v^{(f)}_k(t_1,t_0) \phi_k(t)$,
which satisfy $\Phi^{f\dagger} \Phi^{f} = 1$. Ground and excited state
mass estimates are extracted from effective energies formed from the
correlators
\be
D^{f}(t)=\langle \Phi^{f}(t)^\dagger \Phi^{f}(0) \rangle =
\sum_{k,m} v^{(f)}_k(t_1,t_0) C_{km}(t) v^{(f)}_m(t_1,t_0)
\ee
In some cases we shall be interested in the overlap of our ground
state eigenfunctions onto the basis operators. This is given by
$\langle \phi_k^\dagger \Phi^{f} \rangle = \sum_j C_{kj}(t_0)
v_j^{(f)}(t_1,t_0)$ (which are the left eigenvectors).

\subsection{Non--zero momentum}
\label{sec_non_zero_mom}

We can further improve the mass estimates of the lighter states by
including correlators of non--zero momentum in our analysis. The
lowest such is $\frac{2 \pi}{L}$, and in these units we measure $P
\cdot P = 0,1,2$.  We convert these effective energies to masses using
the lattice dispersion relation
\be 
\hat{E}^{\rm eff}(t;P)^2 = 
\hat{m}^{\rm eff}(t;P)^2 + 
\sum_{\mu=1}^3 \sin^2 \left(\frac{2 \pi P_\mu}{L} \right).
\ee
(This form was found to be more successful than the continuum version
in describing the massless photon dispersion relation in a lattice
Higgs model
\cite{hart_unpub}.)

An immediate concern, however, is that boosting zero momentum irreps
of the cubic group can mix previously orthogonal states. This can
result in contamination of correlation functions by lighter
states. This contamination will increase with the size of the boost
and will particularly involve the lightest, scalar state.

One method of dealing with this is to explicitly subtract off the
scalar contribution. We may determine this by measuring explicitly the
overlap between the operator in question and the boosted scalar
state. Linear combinations of the operators can then render the basis
orthogonal to the scalar contamination. We may get around the problem
more simply, but with the penalty of losing some information and
self--averaging. Consider the continuum notation. We construct
operators with a well defined eigenvalue of a component of $J$, rather
than of $J$ itself. We do this by building a glueball operator with
chosen symmetry under rotation about a particular lattice axis. We
then boost this operator, but only along this axis which does not
change the angular momentum about this axis. We create, in effect, an
eigenstate of helicity. This operator will then couple to the lightest
state with $J$ commensurate to this. For example, if we arrange for
the glueball operator to change by a factor of $-1$ under a rotation
of $\frac{\pi}{2}$ about the $z$--axis, then we fix $J_z=2$. The
operator boosted along the $z$--axis can thus couple only to states
with $J \ge J_z$, which was unaffected by the boost and which excludes
the scalar contamination. The ground state will now be expected to be
purely $J=2$.

In practice, however, the contamination is easy to identify as a
reduction in the effective mass plateaux of the original, boosted
operators. The zero momentum correlation functions and all scalar
states remain, of course, unmixed by the momentum. The $T_2^{++}$
tensor appears unaffected for $P \cdot P = 1$, but does show a reduction
for $P \cdot P = 2$ and we exclude the latter from the mass
analysis. The $E^{++}$ shows a reduction for both non--zero momenta,
and for technical reasons we do not consider this state in this
work. Any possible contamination of higher mass states is within the
large statistical errors of these states.

\begin{figure}[p]
\begin{center}
\leavevmode
\epsfysize=300pt
\epsffile{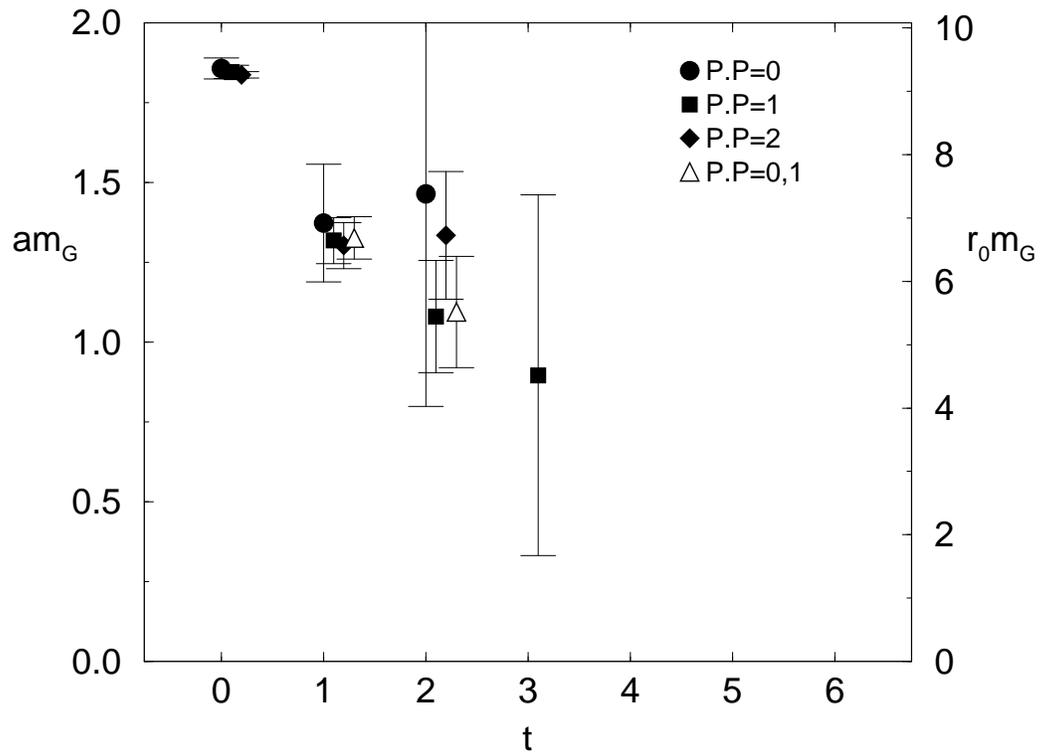}
\end{center}
\caption[]{\label{fig_eff_ex_scalar} {\it 1$^{\rm \it st}$ excited
  $A_1^{++*}$ scalar glueball effective mass plots for the e$_2$ ensemble.}}
\end{figure}

\begin{figure}[p]
\begin{center}
\leavevmode
\epsfysize=300pt
\epsffile{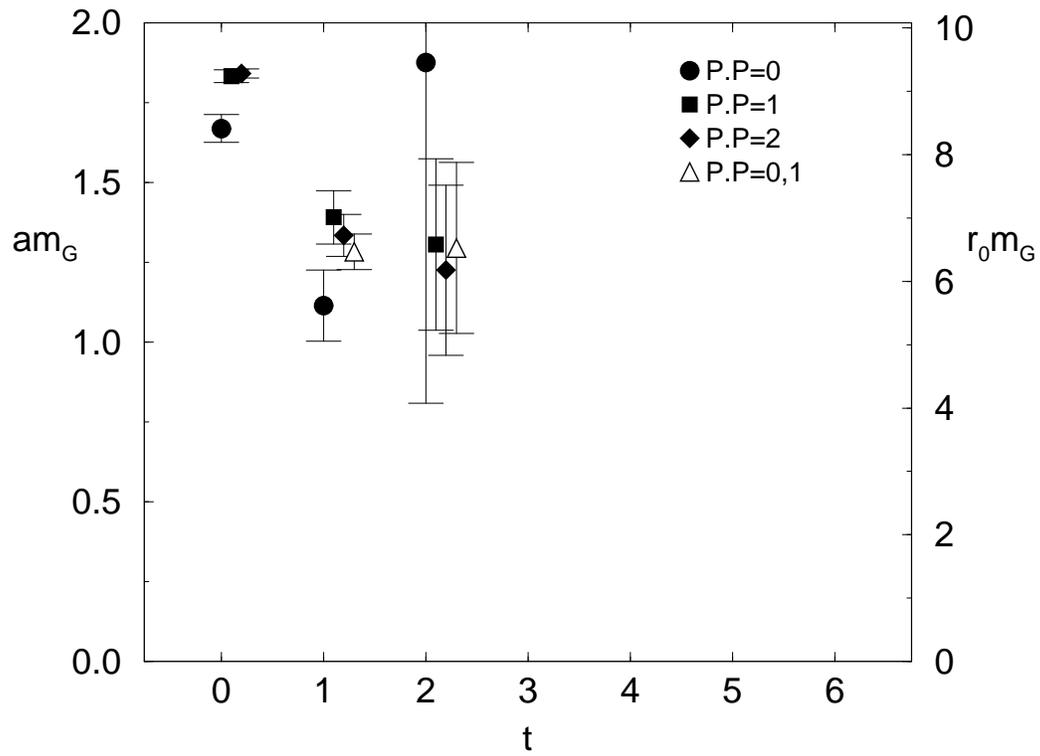}
\end{center}
\caption[]{\label{fig_eff_pseudo} {\it $A_1^{-+}$ pseudoscalar
  glueball effective mass plots for the e$_2$ ensemble.}}
\end{figure}

\begin{figure}[p]
\begin{center}
\leavevmode
\epsfysize=300pt
\epsffile{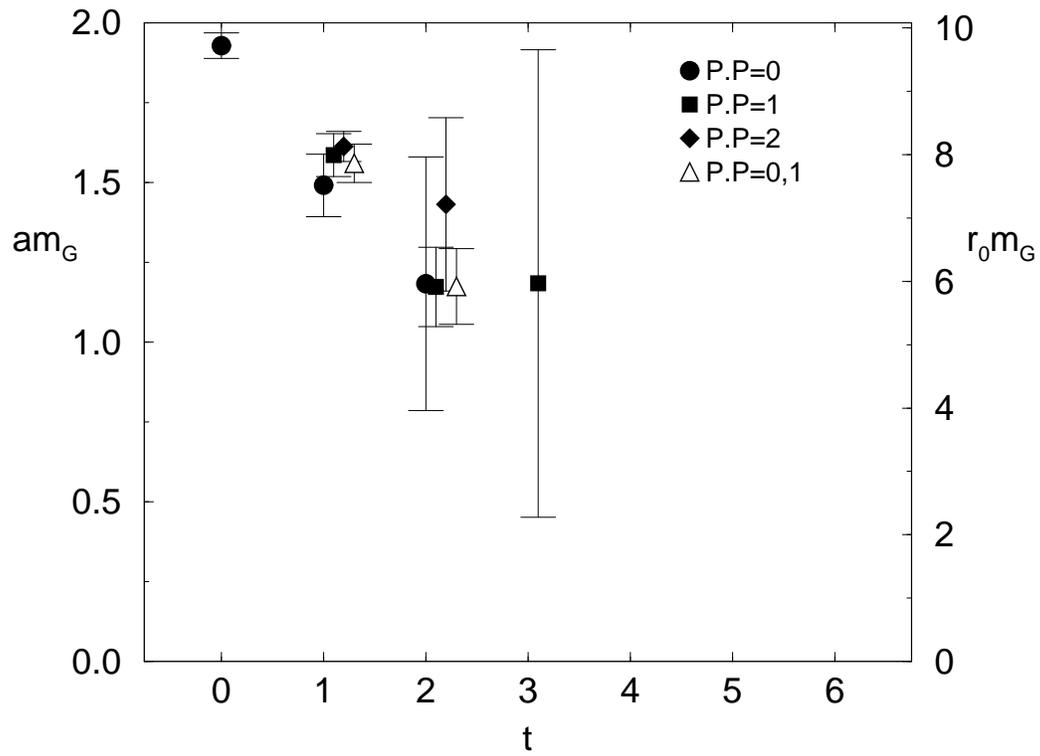}
\end{center}
\caption[]{\label{fig_eff_vector} {\it $T_1^{+-}$ vector glueball
  effective mass plots for the e$_2$ ensemble.}}
\end{figure}

We perform joint fits to the effective mass functions obtained after
variational analysis of correlation matrices and subtraction of the
boost momentum.  For a given state at large $t$, the masses from
differing momenta should become consistent with a single plateau,
which gives the rest mass. The signal from the $P \cdot P=1$ channel
was found to be particularly useful. The mass of the ground state
excitation were still small enough for reliable effective energy
plateaux to be observed for the lighter rest mass states, and
statistical noise was observed to be only of a similar magnitude to
the $P \cdot P=0$ channel. For $P \cdot P=2$, however, the energies of
the states were often too large to be confidently assessed. Where
extractable, the $P \cdot P=2$ correlators showed a consistent
effective mass plateau with lower momentum channels, bar the boost
contamination discussed above. As they did not improve the quality of
the fits, however, we estimate the ground state scalar mass by
performing joint plateau fits to $P \cdot P=0,1$ uncorrelated in both
$t$ and $P$, over as a wide a range of $t$ as the quality of fit
indicators $\chi^2$ per degree of freedom and $Q$ permitted. We give
examples of effective mass plots for the tensor state, the heavier
$1^{\rm st}$ excited scalar, the pseudoscalar and vector states in
Figs.~\ref{fig_eff_ex_scalar}--\ref{fig_eff_vector}. Details of the
fits made are shown in
Tables~\ref{tab_ex_scalar_fits}--\ref{tab_vector_fits}, where an
italic font indicates no masses were extractable beyond $t=1$.

\newpage

\begin{table}[p]
\begin{center}
\begin{tabular}{|c|c|c|c|c|r@{.}l|r@{.}l|r@{.}l|}
\hline \hline
label &
$\beta$ & $\kappa$ & $c_{\rm sw} $ & $N_{\rm traj.}$ & 
\multicolumn{2}{c|}{$\hat{r}_0$} &
\multicolumn{2}{c|}{$\hat{m}_\pi/\hat{m}_\rho$} &
\multicolumn{2}{c|}{$\hat{r}_0 \hat{m}_\pi$} \\
\hline
e$_1$ & 5.20 & 0.13565 & 2.0171 & 2800 & 5&21 (5)   & \none & 
{\em 1}&{\em 386 (63)} \\
\hline
e$_2$ & 5.20 & 0.13550 & 2.0171 & 8300 & 5&041 (40) & 0&$578^{+13}_{-19}$ &
1&480 (22) \\
e$_3$ & 5.25 & 0.13520 & 1.9603 & 8200 & 5&137 (49) & \none &
{\em 1}&{\em 978 (22)} \\
\hline
e$_4$ & 5.20 & 0.13500 & 2.0171 & 7800 & 4&754 (40) & 0&$700^{+12}_{-10}$ &
1&925 (38) \\
e$_5$ & 5.26 & 0.13450 & 1.9497 & 4100 & 4&708 (52) & 0&$783^{+5}_{-5}$ &
2&406 (27) \\
e$_6$ & 5.29 & 0.13400 & 1.9192 & 3900 & 4&754 (40) & 0&$835^{+7}_{-7}$ &
2&776 (32) \\
\hline \hline
\end{tabular}
\caption{ \label{tab_ensembles} {\em The ensembles studied, with
  measurements made every tenth HMC trajectory. Numbers in italics are
  preliminary estimates.}}
\end{center}
\end{table}
\begin{table}[p]
\begin{center}
\begin{tabular}{|c|c|c|c|*{5}{r@{.}l|}}
\hline \hline
ens. & {\small $t_{\rm low}$} & {\small $t_{\rm high}$} & 
{\small $N_{\rm dof}$} &
\multicolumn{2}{c|}{$\chi^2/{\rm dof}$} & 
\multicolumn{2}{c|}{$Q$} &
\multicolumn{2}{c|}{$\hat{m}_t$} &
\multicolumn{2}{c|}{$\surd{\hat{K}}$} &
\multicolumn{2}{c|}{$\hat{r}_0 \surd{\hat{K}}$} \\
\hline
e$_1$ & 
	2 & 4 & 7 & 0&485 & 0&846 & 0&628 (15) & 0&208 (3) & 1&085 (17) \\
e$_2$ & 
	2 & 4 & 5 & 0&218 & 0&955 & 0&765 (41) & 0&228 (6) & 1&149 (32) \\
e$_3$ & 
	2 & 6 & 9 & 1&389 & 0&187 & 0&683 (24) & 0&216 (4) & 1&111 (23) \\
e$_4$ & 
	2 & 5 & 7 & 0&292 & 0&957 & 0&905 (48) & 0&246 (7) & 1&171 (33) \\
e$_5$ & 
	2 & 4 & 6 & 0&511 & 0&801 & 0&794 (54) & 0&232 (8) & 1&091 (40) \\
e$_6$ & 
	2 & 4 & 5 & 0&638 & 0&671 & 0&832 (35) & 0&237 (5) & 1&126 (26) \\
\hline \hline
\end{tabular}
\caption{ \label{tab_toron_fits}
  {\em Fitting the ground state torelon masses and string tensions.}}
\end{center}
\end{table}
\begin{table}[p]
\begin{center}
\begin{tabular}{|c|*{2}{r@{.}l|}}
\hline
\hline
ens. & 
\multicolumn{2}{c|}{$M_{\mu \nu}(t=0)$} &
\multicolumn{2}{c|}{$M_{\tau \phi}(t=0)$} \\
\hline
e$_1$ & 0&0016 (60)    & $-0$&011 (15) \\
e$_2$ & $-0$&0078 (37) & $-0$&014 (11) \\
e$_3$ & $-0$&0070 (132)      & 0&0018 (53) \\
e$_4$ & 0&0018 (37)    & 0&0058 (65) \\
e$_5$ & $-0$&0028 (58) & $-0$&014 (11) \\
e$_6$ & $-0$&0001 (58) & $-0$&003 (6) \\
\hline
\hline
\end{tabular}
\caption{ \label{tab_mixing} {\em Overlaps between orthogonal
torelons and between $A_1^{++}$ torelon combinations and the scalar glueball.}}
\end{center}
\end{table}
\begin{table}[p]
\begin{center}
\begin{tabular}{|c|c|c|c|*{3}{r@{.}l|}}
\hline \hline
ens. & $t_{\rm low}$ & $t_{\rm high}$ & $N_{\rm dof}$ &
\multicolumn{2}{c|}{$\chi^2/{\rm dof}$} & 
\multicolumn{2}{c|}{$Q$} &
\multicolumn{2}{c|}{$\hat{m}_G(A_1^{++})$} \\
\hline
e$_1$ & 2 & 4 & 6 & 0&352 & 0&909 & 0&677 (75) \\
e$_2$ & 2 & 5 & 8 & 0&675 & 0&714 & 0&628 (30) \\
e$_3$ & 2 & 5 & 8 & 0&540 & 0&827 & 0&617 (41) \\
e$_4$ & 2 & 4 & 6 & 0&923 & 0&477 & 0&626 (41) \\
e$_5$ & 2 & 5 & 8 & 0&641 & 0&744 & 0&654 (30) \\
e$_6$ & 2 & 4 & 5 & 1&768 & 0&116 & 0&654 (40) \\
\hline \hline
\end{tabular}
\caption{ \label{tab_scalar_fits}
  {\em Fitting the ground state scalar glueball masses.}}
\end{center}
\end{table}
\begin{table}[p]
\begin{center}
\begin{tabular}{|c|c|c|c|*{3}{r@{.}l|}}
\hline \hline
ens. & $t_{\rm low}$ & $t_{\rm high}$ & $N_{\rm dof}$ &
\multicolumn{2}{c|}{$\chi^2/{\rm dof}$} & 
\multicolumn{2}{c|}{$Q$} &
\multicolumn{2}{c|}{$\hat{m}_G(T_2^{++})$} \\
\hline
e$_1$ & 2 & 3 & 3 & 0&378 & 0&769 & 0&88 (8) \\
e$_2$ & 2 & 3 & 3 & 0&042 & 0&989 & 1&28 (9) \\
e$_3$ & 2 & 3 & 4 & 1&651 & 0&158 & 1&12 (9) \\
e$_4$ & 2 & 3 & 4 & 0&772 & 0&543 & 1&33 (9) \\
e$_5$ & 2 & 3 & 4 & 0&341 & 0&850 & 1&18 (12) \\
e$_6$ & 2 & 3 & 3 & 0&399 & 0&754 & 1&14 (11) \\
\hline \hline
\end{tabular}
\caption{ \label{tab_tensor_fits}
  {\em Fitting the ground state tensor glueball masses.}}
\end{center}
\end{table}
\begin{table}[p]
\begin{center}
\begin{tabular}{|c|c|c|c|*{3}{r@{.}l|}}
\hline \hline
ens. & $t_{\rm low}$ & $t_{\rm high}$ & $N_{\rm dof}$ &
\multicolumn{2}{c|}{$\chi^2/{\rm dof}$} & 
\multicolumn{2}{c|}{$Q$} &
\multicolumn{2}{c|}{$\hat{m}_G(A_1^{++*})$} \\
\hline
e$_1$ & 2 & 2 & 1 & 0&883 & 0&347 & 1&12 (52) \\
e$_2$ & 2 & 2 & 1 & 0&312 & 0&576 & 1&11 (18) \\
e$_3$ & 2 & 2 & 1 & 1&606 & 0&205 & 0&94 (23) \\
e$_4$ & 1 & 1 & 1 & 1&377 & 0&241 & {\it 1}&{\it 31 (7)} \\
e$_5$ & 1 & 1 & 1 & 2&751 & 0&097 & {\it 1}&{\it 32 (10)} \\
e$_6$ & 2 & 2 & 1 & 0&042 & 0&837 & 1&17 (32) \\
\hline \hline
\end{tabular}
\caption{ \label{tab_ex_scalar_fits} {\em Fitting the 1$^{\rm \it st}$
  excited scalar glueball masses. Numbers in italics are upper bounds
  from $t=1$ only.}}
\end{center}
\end{table}
\begin{table}[p]
\begin{center}
\begin{tabular}{|c|c|c|c|*{3}{r@{.}l|}}
\hline \hline
ens. & $t_{\rm low}$ & $t_{\rm high}$ & $N_{\rm dof}$ &
\multicolumn{2}{c|}{$\chi^2/{\rm dof}$} & 
\multicolumn{2}{c|}{$Q$} &
\multicolumn{2}{c|}{$\hat{m}_G(A_1^{-+})$} \\
\hline
e$_1$ & 2 & 2 & 1 & 0&281 & 0&596 & 1&33 (37) \\
e$_2$ & 2 & 2 & 1 & 0&669 & 0&604 & 1&34 (28) \\
e$_3$ & 2 & 2 & 1 & 0&013 & 0&909 & 1&03 (19) \\
e$_4$ & 1 & 1 & 1 & 0&009 & 0&923 & {\it 1}&{\it 47 (6)} \\
e$_5$ & 2 & 2 & 1 & 0&157 & 0&693 & 1&24 (54)\\
e$_6$ & 2 & 2 & 1 & 0&467 & 0&495 & 1&10 (31) \\
\hline \hline
\end{tabular}
\caption{ \label{tab_pseudo_fits} {\em Fitting the ground state
  pseudoscalar glueball masses. Numbers in italics are upper bounds
  from $t=1$ only.}}
\end{center}
\end{table}
\begin{table}[p]
\begin{center}
\begin{tabular}{|c|c|c|c|*{3}{r@{.}l|}}
\hline \hline
ens. & $t_{\rm low}$ & $t_{\rm high}$ & $N_{\rm dof}$ &
\multicolumn{2}{c|}{$\chi^2/{\rm dof}$} & 
\multicolumn{2}{c|}{$Q$} &
\multicolumn{2}{c|}{$\hat{m}_G(T_1^{+-})$} \\
\hline
e$_1$ & 1 & 1 & 1 & 2&829 & 0&093 & {\it 1}&{\it 44 (7)} \\
e$_2$ & 2 & 2 & 1 & 0&001 & 0&982 & 1&17 (12) \\
e$_3$ & 2 & 2 & 1 & 0&023 & 0&878 & 1&38 (21) \\
e$_4$ & 2 & 2 & 1 & 0&383 & 0&536 & 1&68 (22) \\
e$_5$ & 1 & 1 & 1 & 1&005 & 0&316 & {\it 1}&{\it 54 (10)} \\
e$_6$ & 1 & 1 & 1 & 0&010 & 0&921 & {\it 1}&{\it 60 (9)} \\
\hline \hline
\end{tabular}
\caption{ \label{tab_vector_fits} {\em Fitting the ground state vector
  glueball masses. Numbers in italics are upper bounds from $t=1$ only.}}
\end{center}
\end{table}
\begin{table}[p]
\begin{center}
\begin{tabular}{|c|*{2}{r@{.}l|}r@{.}l|}
\hline
\hline
ens. & 
\multicolumn{2}{c|}{$\hat{m}_G / \surd{\hat{K}}$} &
\multicolumn{2}{c|}{$\hat{r}_0 \hat{m}_G(A_1^{++})$} &
\multicolumn{2}{c|}{ratio} \\
\hline
e$_1$ & 3&25 (30) & 3&53 (40) & 0&96 (12) \\
e$_2$ & 2&76 (15) & 3&17 (15) & 0&86 (5)  \\
e$_3$ & 2&86 (21) & 3&17 (22) & 0&86 (7)  \\
e$_4$ & 2&54 (16) & 2&98 (20) & 0&81 (6)  \\
e$_5$ & 2&82 (21) & 3&08 (15) & 0&84 (5)  \\
e$_6$ & 2&76 (14) & 3&11 (20) & 0&85 (6)  \\
\hline
\hline
\end{tabular}
\caption{ \label{tab_scalar_masses} {\em The scalar glueball masses
  and their ratios to the quenched mass, $\hat{r}_0 \hat{m}_G = 3.68 \,
  (8)$ \cite{lucini01,allton01}.}}
\end{center}
\end{table}
\begin{table}[p]
\begin{center}
\begin{tabular}{|c|*{2}{r@{.}l|}r@{.}l|}
\hline
\hline
ens. & 
\multicolumn{2}{c|}{$\hat{m}_G / \surd{\hat{K}}$} &
\multicolumn{2}{c|}{$\hat{r}_0 \hat{m}_G(T_2^{++})$} &
\multicolumn{2}{c|}{ratio} \\
\hline
e$_1$ & 4&26 (38) & 4&62 (41) & 0&79 (8)  \\
e$_2$ & 5&60 (37) & 6&43 (42) & 1&11 (8)  \\
e$_3$ & 5&16 (39) & 5&73 (44) & 0&98 (8)  \\
e$_4$ & 5&39 (29) & 6&32 (40) & 1&09 (8)  \\
e$_5$ & 5&10 (50) & 5&57 (57) & 0&96 (10) \\
e$_6$ & 4&80 (46) & 5&40 (51) & 0&93 (9) \\
\hline
\hline
\end{tabular}
\caption{ \label{tab_tensor_masses} {\em The tensor glueball masses
  and their ratios to the quenched mass, $\hat{r}_0 \hat{m}_G = 5.82 \,
  (14)$ \cite{lucini01,allton01}.}}
\end{center}
\end{table}
\begin{table}[p]
\begin{center}
\begin{tabular}{|c|*{6}{r@{.}l|}}
\hline
\hline
& 
\multicolumn{4}{c|}{$A_1^{++*}$} &
\multicolumn{4}{c|}{$A_1^{-+}$} &
\multicolumn{4}{c|}{$T_1^{+-}$} \\
\cline{2-13}
ens. & 
\multicolumn{2}{c|}{$\hat{r}_0 \hat{m}_G$} &
\multicolumn{2}{c|}{$\hat{m}_G / \surd{\hat{K}}$} &
\multicolumn{2}{c|}{$\hat{r}_0 \hat{m}_G$} &
\multicolumn{2}{c|}{$\hat{m}_G / \surd{\hat{K}}$} &
\multicolumn{2}{c|}{$\hat{r}_0 \hat{m}_G$} &
\multicolumn{2}{c|}{$\hat{m}_G / \surd{\hat{K}}$} \\
\hline
e$_1$ & 
5&9 (2.7) & 5&4 (2.5) & 6&9 (1.9) & 6&4 (1.8) & \none &    \none \\
e$_2$ & 
5&6 (9) & 4&85 (84) & 6&8 (1.4) & 5&9 (1.3) & 5&92 (60) & 5&15 (53) \\
e$_3$ & 
4&8 (1.2) & 4&4 (1.1) & 5&3 (1.0) & 4&8 (9) & 7&1 (1.1) & 6&4 (1.0) \\
e$_4$ & 
\none &    \none &    \none &    \none &    8&0 (1.0) & 6&83 (93) \\
e$_5$ & 
\none &    \none &    5&9 (2.6) & 5&4 (2.4) & \none &    \none \\
e$_6$ & 
5&6 (1.5) & 5&0 (1.4) & 5&2 (1.5) & 4&7 (1.4) & \none &    \none \\
\hline
\hline
\end{tabular}
\caption{ \label{tab_other_masses} {\em Various other glueball masses,
including only estimates from $t \ge 2$.}}
\end{center}
\end{table}
\begin{table}[p]
\begin{center}
\begin{tabular}{|c|r@{.}l|r@{.}l|}
\hline
\hline
\multicolumn{5}{|c|}{$\hat{m}^{\rm eff}(t)$}  \\
\hline
$t$ &  
\multicolumn{2}{c|}{dynamical}  &
\multicolumn{2}{c|}{quenched} \\ 
\hline
0 & 0&851 (23) & 0&901 (6) \\
1 & 0&638 (24) & 0&795 (11) \\
2 & 0&580 (41) & 0&754 (28) \\
3 & 0&768 (80) & 0&740 (49) \\
\hline \hline
\end{tabular}
\caption{ \label{tab_statistics} {\em The scalar glueball effective
mass as a function of $t$, comparing the $P \cdot P=0$ values from the
$e_2$ ensemble with quenched calculations at $\beta=5.93$.}} 
\end{center}
\end{table}

\clearpage

\begin{figure}[p]
\begin{center}
\leavevmode
\epsfysize=300pt
\epsffile{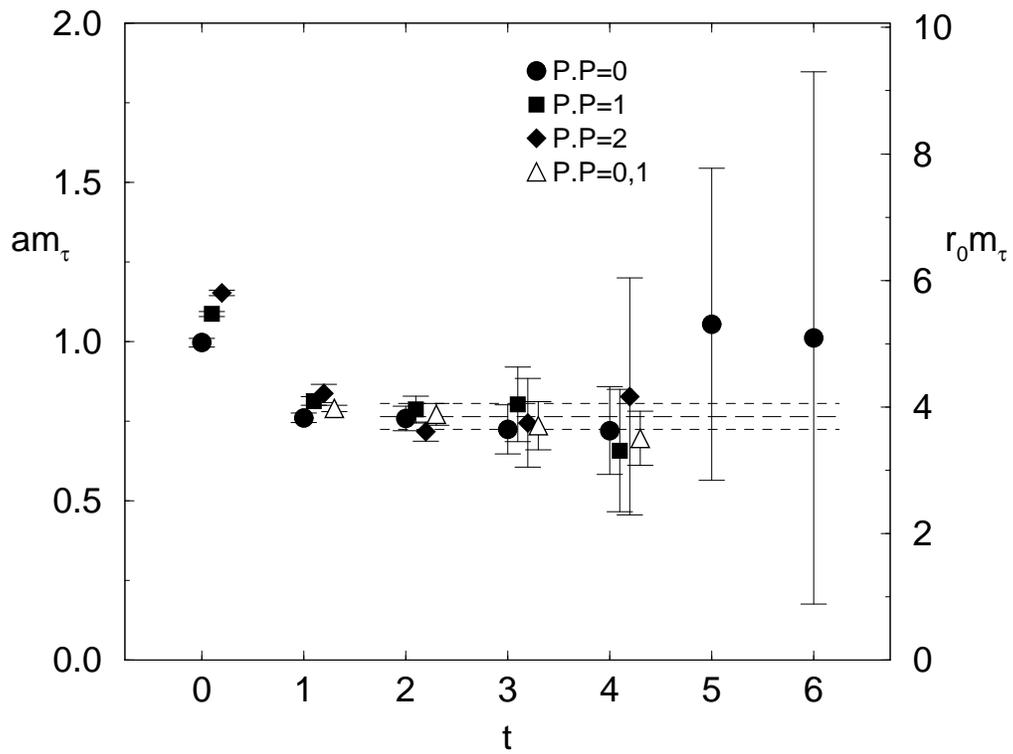}
\end{center}
\caption[]{\label{fig_eff_toron} {\it Torelon effective mass plot for
the e$_2$ ensemble, with fitted plateau indicated.}}
\end{figure}
\begin{figure}[p]
\begin{center}
\leavevmode
\epsfysize=300pt
\epsffile{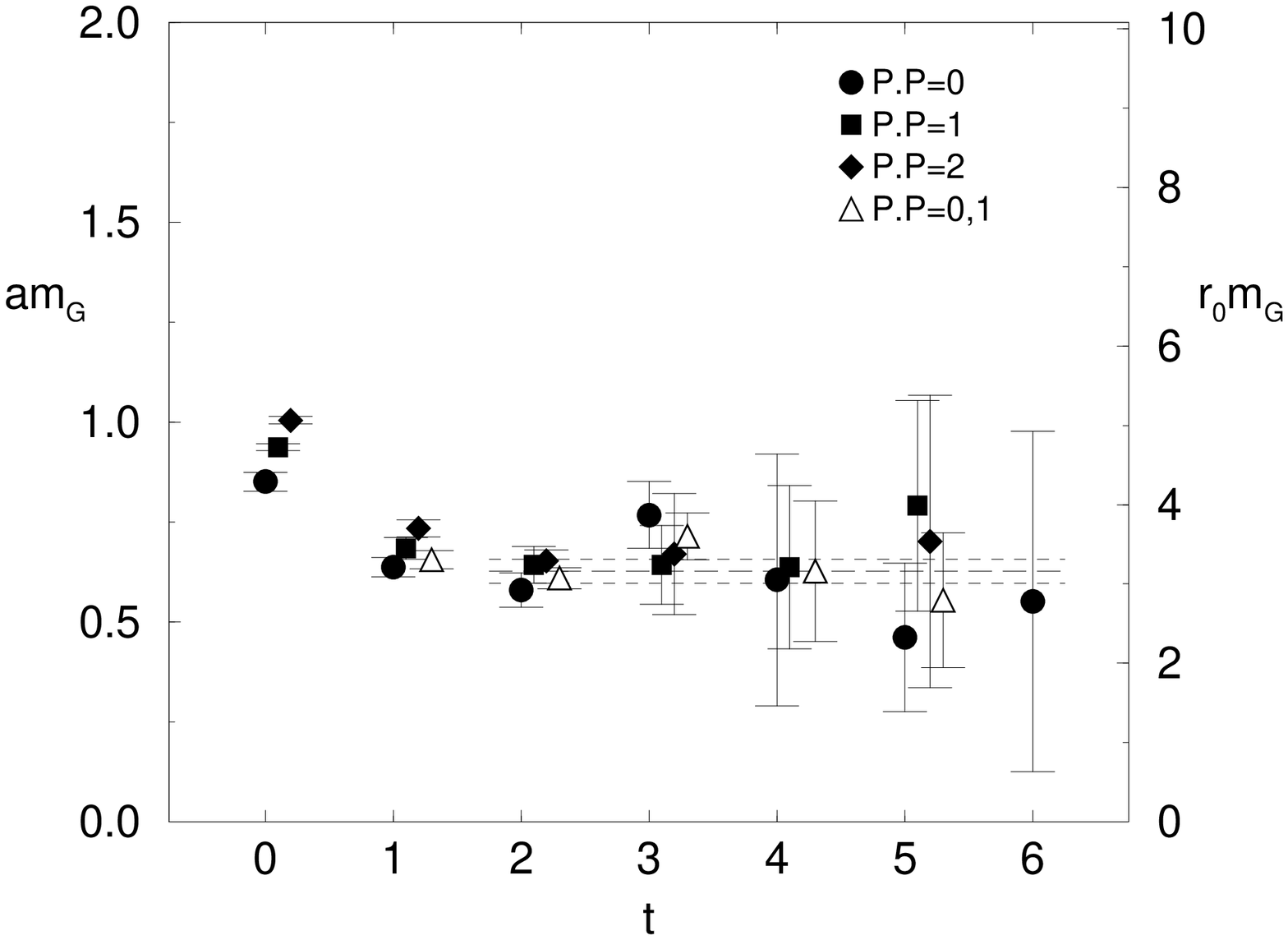}
\end{center}
\caption[]{\label{fig_eff_scalar} {\it Scalar glueball effective mass
  plots for the e$_2$ ensemble, with fitted plateau
  indicated.}}
\end{figure}
\begin{figure}[p]
\begin{center}
\leavevmode
\epsfysize=300pt
\epsffile{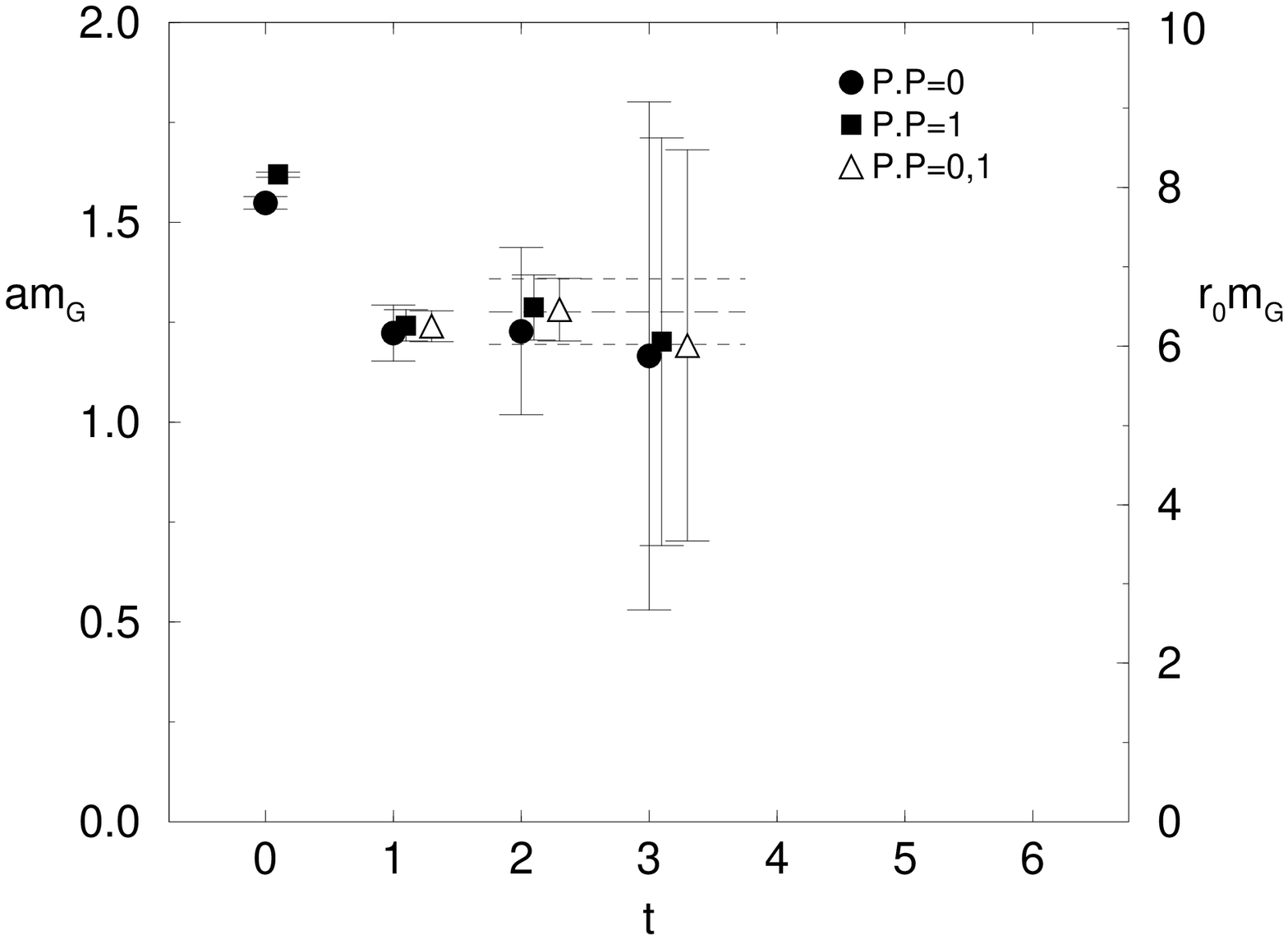}
\end{center}
\caption[]{\label{fig_eff_tensor} {\it Tensor glueball effective mass
  plots for the e$_2$ ensemble, with fitted plateau
  indicated.}}
\end{figure}

\end{document}